\documentclass[prb,twocolumn,aps,superscriptaddress,10pt]{revtex4-2}
\usepackage{graphicx}
\usepackage{dcolumn}
\usepackage{amsmath}
\usepackage{amssymb}
\usepackage{bm}
\usepackage{mathtools}
\usepackage{physics}
\usepackage[utf8]{inputenc}
\usepackage[usenames, dvipsnames]{color}
\usepackage[english]{babel}
\usepackage[T1]{fontenc}
\usepackage{bbm}
\usepackage{float}
\usepackage{verbatim}
\usepackage[caption=false]{subfig}
\usepackage{xfrac}
\usepackage{upgreek}
\usepackage{makecell}
\usepackage{comment}
\usepackage{placeins}
\usepackage[normalem]{ulem}
\usepackage{hyperref}
\usepackage{xcolor}
\hypersetup{colorlinks = true, linkcolor=blue, citecolor=blue, urlcolor=blue}

  \makeatletter
    \renewcommand\@make@capt@title[2]{%
     \@ifx@empty\float@link{\@firstofone}{\expandafter\href\expandafter{\float@link}}%
      {\textsc{#1}}\@caption@fignum@sep#2\quad}%
    \makeatother

\begin{document}
\title{Current cross-correlations as probes for poor man's Majorana states}

\author{Saatwik Patnaik}
\affiliation{Department of Electrical Engineering, Indian Institute of Technology Bombay, Powai, Mumbai-400076, India}
\author{Aditya Saran}
\affiliation{Department of Physics, Indian Institute of Technology Bombay, Powai, Mumbai-400076, India}
\author{Himadri S Dhar}
\affiliation{Department of Physics, Indian Institute of Technology Bombay, Powai, Mumbai-400076, India}
\author{Pertti~Hakonen}
\affiliation{Low Temperature Laboratory, Department of Applied Physics, Aalto University, PO Box 15100, FI-00076, Espoo, Finland}
\affiliation{InstituteQ, Department of Applied Physics, Aalto University, PO Box 15100, FI-00076, Espoo, Finland
}
\author{Thierry Martin}
\affiliation{Aix Marseille Univ, Université de Toulon, CNRS, CPT, Marseille, France}
\author{Bhaskaran Muralidharan}
\email{Corresponding author: bm@ee.iitb.ac.in}
\affiliation{Department of Electrical Engineering, Indian Institute of Technology Bombay, Powai, Mumbai-400076, India} 
\date{\today}
\begin{abstract} 
 The minimal Kitaev chain that emulates a topological superconductor with three quantum dots offers a tunable platform for potentially hosting poor man's Majorana (PMM) modes. 
 Asserting the need to go beyond differential conductance spectroscopy, we introduce current-current correlations as a viable framework for verifying their true non-locality. The robustness of the PMM modes, specifically with respect to delocalization as the system is tuned away from {\it{sweet spots}}, we show, is embedded in the relative magnitudes of the nonlocal transport processes. This aspect is adeptly captured by current cross-correlations, whose features show remarkable stability around the PMM sweet spot, specifically with respect to the detuning of an outer dot.  We establish this as a prominent feature and a diagnostic for true PMMs even in the short chain limit. Our results accentuate the need for current cross-correlation measurements as a diagnostic framework for unambiguously verifying true non-locality of entangled states as well as topologically protected states.
\end{abstract}
\maketitle
\section{Introduction}
\noindent
\begin{figure*}[hbt]
\centering
 \includegraphics[width=2\columnwidth]{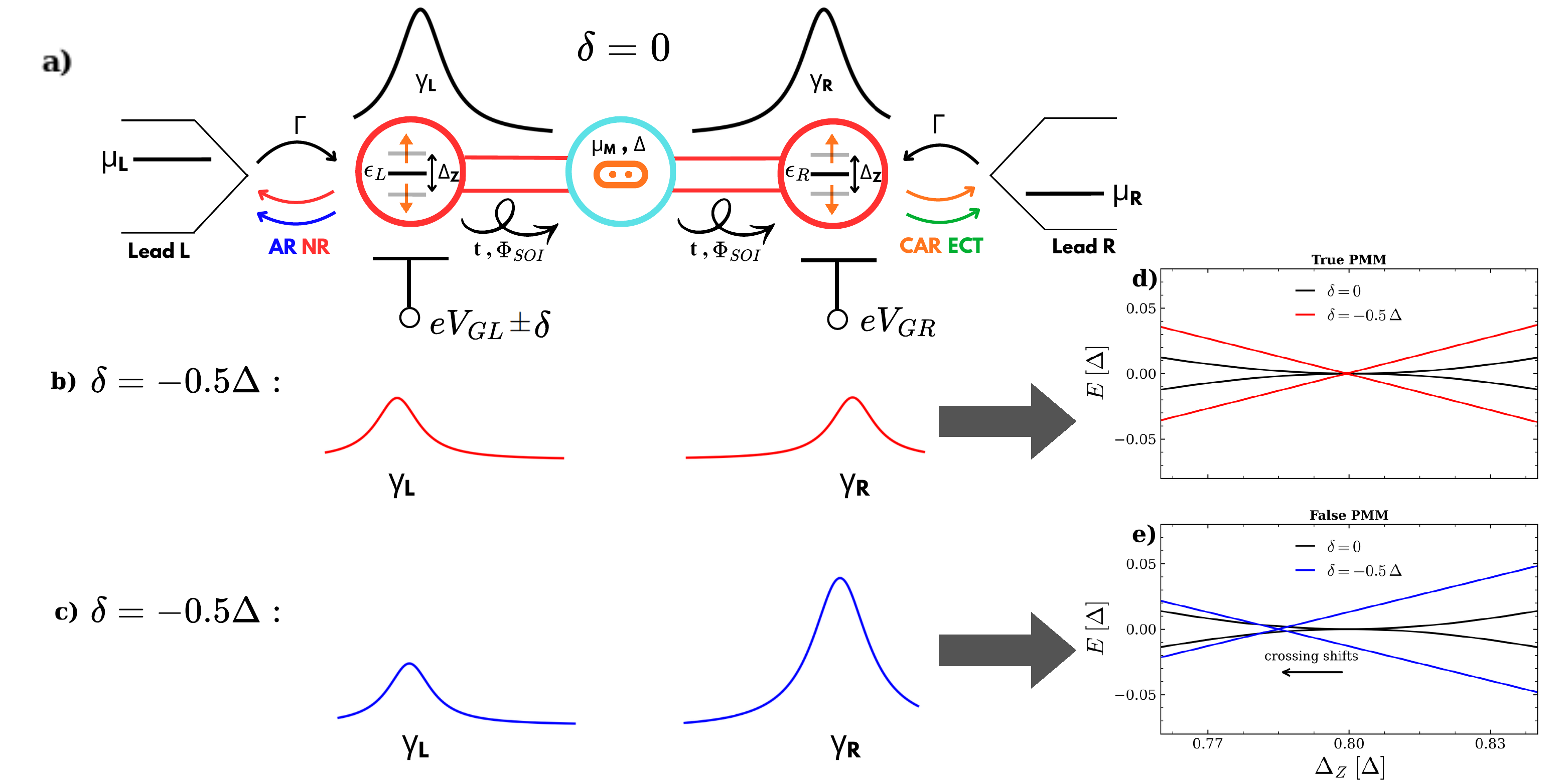}
  \caption{Preliminaries. (a) Device schematic of the MKC with three quantum dots. The middle dot is superconducting, and the outer dots have spin-split energy levels due to the presence of an external magnetic field. The energy levels of the outer dots can be controlled via a gate potential. The interdot tunneling has both spin-conserving and spin-flip tunnel processes facilitated by the Rashba spin-orbit interaction. 
  The MKC is coupled to normal metallic leads through which currents flow. At the interfaces between the leads and the MKC system, one has four local (nonlocal) processes, namely, the NR (ECT) and the AR (CAR) processes. (b) A schematic of a true PMM mode that does not hybridize upon detuning an outer dot (say QDL), (c) a schematic of the false PMM mode that hybridizes upon detuning of the dot. d) and e) The corresponding energy landscapes around the sweet spots for outer dot detunings $\delta =0$ and $\delta =-0.5 \Delta$. }
\label{fig:Fig1}
\end{figure*}
 The search for Majorana quasiparticles \cite{kitaev2001unpaired,liu2015probing, lee2012zero, deng2016majorana, liu2017andreev, moore2018quantized, woods2019zero, prada2020andreev,yu2021non, wang2023triplet} in condensed matter systems 
is an extensively pursued goal not only due to interesting fundamental physics, but also as building blocks of topologically protected quantum computation \cite{Topo_QC_1,Marra}. 
The proposal to use semiconductor nanowires with strong spin-orbit coupling proximitized to a s-wave superconductor \cite{Lutchyn,Oreg,Marra} motivated several experimental explorations \cite{van2026single,liu2015probing, lee2012zero, deng2016majorana, liu2017andreev, moore2018quantized, woods2019zero, prada2020andreev,yu2021non, wang2023triplet} in hybrid semiconductor-BCS superconductor devices. 
However, material disorder and the presence of unwanted boundary effects have proven very challenging to address, as these can sometimes mimic the experimental signatures of true Majorana zero modes (MZMs). These experimental signatures are typically probed via conductance spectroscopy. In the current context, it is important to explore diagnostic methods beyond conductance spectroscopy \cite{Flindt_2,Arijit,Smirnoff_1} to delineate the true nonlocality and topological order \cite{Kejri,Arora_2024,Ojanen}. In this paper, we provide key insights into a quantum noise (current-current correlations) diagnosis \cite{Flindt_2,Arijit,Smirnoff_1} aimed at distinguishing the signatures of true nonlocality.\\
\indent The paradigm of poor man's Majorana (PMM) \cite{leijnse2012parity,sau2012realizing}, which aims to emulate the Kitaev model \cite{kitaev2001unpaired} with an alternating series of normal quantum dots juxtaposed with superconducting segments -- a geometry proposed in Ref. \cite{sauret_PRB04} to implement a teleportation protocol, has gained a lot of attraction \cite{dvir2023realization, tsintzis2022creating,ten2024two, fulga2013adaptive,li2014tunable, dvir2023realization, liu2023fusion, souto2023probing, samuelson2024minimal, bordin2024crossed, zatelli2024robust, ten2025observation,Klin_1,luethi2025fate,bordin2025enhanced, zhang2026gate, bordin2024signatures, van2024cross,Yeyati_2}. Specifically, the minimal Kitaev chain (MKC) setup sketched in Fig.~\ref{fig:Fig1}(a), also resembles the Cooper pair splitter (CPS) configuration, which has been very well studied, both theoretically and experimentally, even in the context of solid-state entanglement generation \cite{electronicentanglement1,electronicentanglement2,chevallier2011current,Flindt_1,Flindt_2,Arnav,CPS_Hakonen,CPS_Hakonen_2,CPS_Leo}.  They are not true MZMs in the sense of true topological protection due to the absence of a proper bulk. Despite this, the possibility of manipulating the PMM modes for non-Abelian braiding and fusion \cite{PRXQuantum.5.010323} has also been proposed. \\
\indent In MKC, crossed Andreev processes \cite{Byers_PRL1995,anantram1996current,Martin1996,LesovikEPJB2001,Chtchelkatchev2002} give rise to an effective superconducting pairing between two normal dots, and the electron co-tunneling through the superconductor plays the role of hopping between them.  A toy model for PMM can be constructed from minimal spinless models \cite{leijnse2012parity,prada2020andreev,Souto2024}. A simplistic viewpoint that emerges from this toy model is that a PMM operating point occurs when the crossed Andreev process and the elastic cotunneling process balance each other, which happens at zero energy. However, the experimental realization necessitates spin-conserving and spin-non-conserving tunneling processes, resulting from the interplay of Rashba spin-orbit coupling and an applied Zeeman field. \\
\indent Zero energy PMM modes emerging in these finite-sized chains have also been experimentally realized \cite{dvir2023realization, tsintzis2022creating,ten2024two, fulga2013adaptive,li2014tunable, dvir2023realization, liu2023fusion, souto2023probing, samuelson2024minimal, bordin2024crossed, zatelli2024robust, ten2025observation, bordin2025enhanced, zhang2026gate, bordin2024signatures, van2024cross} via conductance spectroscopy.  More recently, PMMs in an MKC were probed for their non-locality via single-shot readouts of the quantum capacitance \cite{parity_readout}. However, PMM states, similar to MZMs \cite{leijnse2012parity, zou2026charge, tsintzis2022creating, zatelli2024robust}, exist only at fine-tuned points in the parameter space, called {\it{sweet spots}}. Even at identified sweet spots, as shown in Ref. \cite{luethi2025fate}, trivial zero-energy states can mimic many characteristics of true PMMs. These were then distinguished by theoretically extending to an infinite chain, keeping the same parameters, where a topological regime is observed in the true PMM case, unlike the trivial (false) case. It was also established \cite{luethi2025fate} that conductance spectroscopy is inadequate in terms of making this distinction. This limitation motivates the search for alternative, experimentally accessible probes that can resolve the topological character of PMMs without requiring a theoretical extension to the infinite-chain limit. \\
\indent  Indeed, in the prior investigation of ``long'' topological superconductor experiments, when the first experimental evidence of a zero bias peak in the differential resistance \cite{mourikSCIENCE2012} was discovered, efforts were made to develop quantum noise based detection protocols \cite{jonckheerePRB2017,jacquetEPJB2018,bathelierPRB2019,jonckheerePRL2019}, using the low energy theory of Ref. \cite{ZazunovPRB2016} to distinguish true signatures of topological superconductivity from other systems with similar differential conductance features.
 A critical signature of nonlocal spin-singlet-type quantum correlations present in such setups is the positive sign of cross-correlations \cite{anantram1996current,electronicentanglement1,CPS_Keldysh4,Burkard_Loss}. \\ 
\indent Here, we further the exploration by detuning an outer quantum dot away from the sweet spot regime. We demonstrate distinct transport physics in the case of true and false PMMs. This expectation also stems from the fact that true PMMs connect to the topological regime in the long chain limit \cite{luethi2025fate}, and thus their behavior in the short chain should be similarly different from that of a false PMM under detuning. To exemplify our point, we show from Figs.~\ref{fig:Fig1}(b) and (c) that this stability against detuning depends on whether one has a true or a false PMM, whose near-identical spectra are shown in Figs.~\ref{fig:Fig1} (d) and (e), respectively. Our analysis reveals that the relative weights of the elastic cotunneling (ECT) and crossed Andreev reflection (CAR) transmissions can track the underlying Majorana wavefunctions and their hybridization under outer dot detuning. \\
\indent We find that current correlations and particularly cross-correlations \cite{Samuel} provide a deeper insight into this problem by encoding the two-particle interference of the quasiparticles propagating between the two leads. We establish that this subtlety, which is captured by the cross-correlation as a sign change at the sweet spot, reflects the necessary stability of the true PMM modes under detuning. Indeed, the delicate balance between the CAR and ECT processes, and their connection with PMM delocalization under detuning (Figs.~\ref{fig:Fig1}(b) and (c)), is embedded in the cross-correlation signatures that not only survive but also get amplified with applied bias.  \\
\indent This paper is structured as follows. In Sec. \ref{sec:formula}, we introduce the device Hamiltonian and formalism for quantum transport, with results for different types of transmission that are possible through the device. In Sec. \ref{sec:results}, we illustrate the results of the current correlations with the transport processes.  We then focus on the stability of PMMs on detuning the left quantum dot in Sec. \ref{sec:detune}, followed by detailed discussions on the results in Sec. \ref{disc}.  We summarize the work with a brief outlook in Sec. \ref{sec:conclusion}.
\section{Formulation}\label{sec:formula}
The transport setup is schematized in Fig.~\ref{fig:Fig1} (a) and comprises the device region along with the two contacts/leads. The device region comprises the MKC with three dots, where a central superconducting dot represented by an Andreev bound state (ABS) is flanked by two normal quantum dots. The energy levels of the two dots can be varied with the local gate electrodes. The left quantum dot (QDL) and the right quantum dot (QDR) are connected to the left and right contacts (electrodes) to facilitate transport spectroscopy measurements. The contacts are held at their electrochemical potentials $\mu_L$ and $\mu_R$. This configuration allows for independent tuning of the left- and right-dot energies, enabling access to the sweet-spot condition required for PMM formation.
\subsection{Model Hamiltonian}
We consider a system of three coupled quantum dots, as in Fig \ref{fig:Fig1}(a), for the MKC. We closely follow the model Hamiltonian from Ref. \cite{luethi2025fate} to find established sweet spots to utilize for our in-depth quantum transport analysis. The Hamiltonian for our device region hence reads
\begin{equation}\label{MKC}
    \begin{aligned}
    &\hat{H} =\sum_{j=L,R}\left[\sum_{\sigma=\uparrow,\downarrow}(\hat{\epsilon}_{j} + \sigma\Delta_Z)\,\hat{n}_{j,\sigma} \right] \\[4pt]
    &+ \sum_{\sigma=\uparrow,\downarrow}\epsilon_M\, \hat{c}_\sigma^\dagger \hat{c}_\sigma+ \Delta\,(\hat{c}_\uparrow^\dagger \hat{c}_\downarrow^\dagger + \hat{c}_\downarrow \hat{c}_\uparrow) \\[4pt]
    &+ t\sum_{\sigma,\sigma'=\uparrow,\downarrow}\left[\left(U_{\mathrm{SOI}}\!\left(\tfrac{\Phi_{\mathrm{SOI}}}{2}\right)\right)_{\!\sigma\sigma'}\!\!\bigl(\hat{c}_\sigma^\dagger \hat{d}_{L,\sigma'} + \hat{d}_{R,\sigma'}^\dagger \hat{c}_\sigma\bigr) +\mathrm{h.c.}\right],
    \end{aligned}
\end{equation}
where the onsite energies are $\hat{\epsilon_L} = \epsilon_L +eV_{GL} \pm \delta$, $\hat{\epsilon_R} = \epsilon_L +eV_{GR} $, denoting the electrostatic influence of the individual gates connected to either dot. Gate voltages $V_{GL(GR)}$ are effective potentials, after taking into account the effective electrostatic environment. The operator $\hat{d}_{j,\sigma}$ ($\hat{d}_{j,\sigma}^\dagger$) annihilates (creates) an
electron with spin $\sigma$ on the left ($j=L$) or right ($j=R$)
quantum dot, while $\hat{c}_\sigma$ ($\hat{c}_\sigma^\dagger$) acts on the middle superconducting dot. The corresponding number operators are
$\hat{n}_{j,\sigma}=\hat{d}_{j,\sigma}^\dagger \hat{d}_{j,\sigma}$.
The parameters $\epsilon_j$ and $\epsilon_M$ denote the on-site energies
of the outer and middle dots, respectively, and $\Delta_Z$ is the Zeeman splitting generated by an external magnetic field. \\
\indent The middle dot is proximity coupled to a superconductor, which gives
rise to an induced pairing amplitude $\Delta$.
The hopping amplitude between neighboring dots is denoted by $t$, while
$U_{\mathrm{SOI}}$ describes spin rotations arising from spin-orbit
interaction, parameterized by the spin-orbit angle
$\Phi_{\mathrm{SOI}}$. The spin-orbit interaction matrix is
\begin{equation}
U_{\mathrm{SOI}}\!\left(\frac{\Phi_{\mathrm{SOI}}}{2}\right) \;=\; \cos\!\left(\frac{\Phi_{\mathrm{SOI}}}{2}\right)I \;+\; i\sin\!\left(\frac{\Phi_{\mathrm{SOI}}}{2}\right)\sigma_y,
\end{equation}
The full Hamiltonian on the $M$ coupled quantum dots in the BdG space reads
\begin{equation}
  \hat{H}_{BdG} \;=\; \frac{1}{2}\left(\sum_{j=1}^{M} \hat{\Psi}_j^{\dagger}\,\alpha_j\,\hat{\Psi}_j
        \;+\; \sum_{j=1}^{M-1}\Bigl(\hat{\Psi}_{j+1}^{\dagger}\,\beta\,\hat{\Psi}_j \;+\; \mathrm{h.c.}\Bigr)\right ),
        \label{eq:Nambu}
\end{equation}
where the on-site and hopping blocks are
\begin{align}
  \alpha_j &\;=\; \bigl(\epsilon_j\,\sigma_0 + \delta_{Z,j}\,\sigma_z\bigr)\otimes\tau_z
                \;+\; \Delta_j\,\sigma_y\otimes\tau_y , \\[4pt]
  \beta    &\;=\; t\,\Bigl(\cos\tfrac{\Phi_{\mathrm{SOI}}}{2}\,\sigma_0
                          + i\sin\tfrac{\Phi_{\mathrm{SOI}}}{2}\,\sigma_y\Bigr)\otimes\tau_z .
\end{align}  
In the BdG representation, $\Psi_j$ denotes the Nambu spinor at the site $j$,
defined as $\hat{\Psi}_j =\begin{pmatrix}
\hat{d}_{j,\uparrow} 
\hat{d}_{j,\downarrow} 
\hat{d}_{j,\uparrow}^\dagger
\hat{d}_{j,\downarrow}^\dagger
\end{pmatrix}^T$, where $T$ stands for transpose.
The matrices $\sigma_i$ and $\tau_i$ ($i=x,y,z$) are Pauli matrices that
act on the spin and particle-hole spaces, respectively, while
$\sigma_0$ denotes the $2\times2$ identity matrix.
The parameter $\delta_{Z,j}$ represents the site-dependent Zeeman field and $\Delta_j$ the
induced superconducting pairing potential. \\
\indent The MKC Hamiltonian in \eqref{MKC} can be used to find several sweet spots by setting thresholds on the purity metrics of PMM \cite{tsintzis2022creating,luethi2025fate}. Furthermore, as elucidated in \cite{luethi2025fate}, the sweet spots may be further classified as true and false PMMs depending on how the states evolve in the long chain limit. We start with well established sweet spots as described in Ref. \cite{luethi2025fate}, using $\Delta_z = 0.8\Delta$, with the dot energy levels tuned as described in Tab.~\ref{tab:model-params}. We show in Figs.~\ref{fig:Fig1} (b) and (c) the schematic of the PMM states under left dot detuning with the anticipation that true PMMs do not delocalize much under these circumstances. Shown in Figs.~\ref{fig:Fig1} (d) and (e) are the near-identical energy spectra with Zeeman energy $\Delta_z$ for the true and false PMM cases, respectively. We now focus on the transport signatures around these sweet spots, simulating the model Hamiltonian using the parameter values listed in Table \ref{tab:model-params}, following Ref. \cite{luethi2025fate}.
\begin{table}[h]
\caption{Model Hamiltonian parameters for the true and false PMM sweet spots \cite{luethi2025fate}.
$t$: interdot hopping; $\Phi_{\rm SOI}$: spin-orbit angle;
$\epsilon_{L,R}$: outer-dot on-site energy at the sweet spot
($\epsilon_L = \epsilon_R$); $\epsilon_M$: middle dot on-site energy;
$U = 0$ throughout (enabling the BdG reduction). Both sweet spots occur at
$\Delta_Z^{\rm sw} = 0.8\,\Delta$. All energies in units of $\Delta$.}
\label{tab:model-params}
\begin{ruledtabular}
\begin{tabular}{lcc}
 & True PMM & False PMM \\
\hline
$t\;[\Delta]$                 & $0.42$    & $0.99$    \\
$\Phi_{\rm SOI}\;[\rm rad]$   & $0.26\pi$ & $0.44\pi$ \\
$\epsilon_{L,R}\;[\Delta]$    & $0.884$   & $0.593$   \\
$\epsilon_M\;[\Delta]$        & $1.275$   & $-3.836$  \\
$\Delta_Z^{\rm sw}\;[\Delta]$ & $0.8$     & $0.8$     \\
\end{tabular}
\end{ruledtabular}
\end{table}

\subsection{Transport theory}
To understand quantum transport through such a device, we use the scattering matrix \cite{Martin1996,buttiker1992scattering, anantram1996current,Martin2005,Torres2001} formalism to obtain closed-form expressions for the currents and current correlation signatures. One then needs to calculate the individual $s$-matrix elements that can be defined with respect to the electron-hole sectors, as described in detail in App. \ref{smatrix}. Using this information, we can extract the reflection and transmission operators, of size 2 $\times$ 2, by calculating the magnitudes of the scattering matrix elements.
\begin{eqnarray}
    \hat{R}_N(E) = s^{\dagger}_{Le,Le}(E)s_{Le,Le}(E) \nonumber  \\
    \hat{R}_A(E) = s^{\dagger}_{Lh,Le}(E)s_{Lh,Le}(E) \nonumber \\
    \hat{T}_{ECT}(E) =s^{\dagger}_{Re,Le}(E)s_{Re,Le}(E) \nonumber \\ 
    \hat{T}_{CAR}(E) =s^{\dagger}_{Rh,Le}(E) s_{Rh,Le}(E)
    \label{tran},
\end{eqnarray}
where $R_{N}$ denotes the normal reflection (NR), $R_A$ denotes the Andreev reflection (AR), and $T_{ECT(CAR)}$ represents the transmission due to the ECT (CAR) process. 
The $s$-matrix convention here is $s_{i \alpha,j \beta} = s_{i \alpha \leftarrow j \beta}$, representing the transition amplitude from the $j \beta$ contact to the $i \alpha$ contact, where $i \in (L,R)$, $j \in (L,R)$ , $\alpha \in (e,h)$ and $\beta \in (e,h)$. 
\subsection{Conductance and Current Correlations}
\label{sec:shotnoise}
Using the reflection and transmission operators defined above, we can now define the overall reflection and transmission for the multi-mode channel by taking a trace as follows:
\begin{eqnarray}
    {R}_N(E) = \mathbf{Tr} [s^{\dagger}_{Le,Le}(E)s_{Le,Le}(E)] \nonumber  \\
    {R}_A(E) =\mathbf{Tr}[s^{\dagger}_{Lh,Le}(E) s_{Lh,Le}(E)] \nonumber \\
    {T}_{ECT}(E) =\mathbf{Tr}[s^{\dagger}_{Re,Le}(E)s_{Re,Le}(E)] \nonumber \\ 
    {T}_{CAR}(E) = \mathbf{Tr}[s^{\dagger}_{Rh,Le}(E) s_{Rh,Le}(E)]
    \label{tran2}.
\end{eqnarray}
First, we must note that the Hamiltonian described in the previous section describes the MKC system. The external leads are connected to QDL and QDR, with a net applied bias $V$ given by $eV= \mu_L-\mu_R = e(V_L - V_R)$, corresponding to a voltage $V_{L(R)}$ at either end. The above can then be used to calculate the terminal electronic current at the left contact, for example, within the Landauer B\"{u}ttiker form:
\begin{equation} \label{2}
\begin{aligned}
I_{L}^{(e)}=&-\frac{e}{h} \left\{\int d E R_{A}(E)\left[f\left(E-e V_{L}\right)-f\left(E+e V_{L}\right)\right]\right.\\
&+\int d E T_{C A R}(E)\left[f\left(E-e V_{L}\right)-f\left(E+e V_{R}\right)\right] \\
&\left.+\int d E T_{ECT}(E)\left[f\left(E-e V_{L}\right)-f\left(E-e V_{R}\right)\right]\right\}.
\end{aligned}
\end{equation}
In this work, we apply a bias in one contact and $V_{L(R)}$ represents the bias applied in the left (right) lead with $\mu_{L(R)} = eV$. Note that the expression above is evaluated in the Nambu electron-hole space, where each contact $L(R)$ is subdivided into two contacts $Le(Re), Lh(Rh)$ \cite{Kejri,MURALIDHARAN202661}. The respective electrochemical potentials are $\mu_{Le (Re)}= eV_{L(R)}$ and $\mu_{Lh(Rh)}=-eV_{L(R)}$.
\\
\indent With the above definition of the current operator and based on the bias situation, the expression for the conductance matrix $[G]$ can be easily derive as follows:
\begin{equation} \label{eq3}
\mathrm{[G]}=\left(\begin{array}{cc}
G_{L L} & G_{L R} \\
G_{R L} & G_{R R}
\end{array}\right)=\left(\begin{array}{ll}
\left.\frac{\partial I_{L}}{\partial V_{L}}\right|_{V_{R}=0} & \left.\frac{\partial I_{L}}{\partial V_{R}}\right|_{V_{L}=0} \\
\left.\frac{\partial I_{R}}{\partial V_{L}}\right|_{V_{R}=0} & \left.\frac{\partial I_{R}}{\partial V_{R}}\right|_{V_{L}=0}
\end{array}\right),
\end{equation}
Taking the zero-temperature limit, $T \rightarrow 0$, of the above equation yields
\begin{equation} \label{4}
\begin{aligned}
\left.G_{LL}(V)\right|_{T \rightarrow 0} \equiv \frac{e^2}{h}\left[R_{A}(E=e V)+R_{A}(E=-e V)\right.+\\
\left.T_{C A R}(E=eV)+T_{ECT}(E=e V)\right],
\end{aligned}
\end{equation}
for the local conductance, and 
\begin{equation} \label{eq5}
\begin{aligned} 
\left.G_{LR}(V)\right|_{T \rightarrow 0} \equiv \frac{e^2}{h}\left[T_{ECT}(E=e V)-T_{CAR}(E=-e V)\right],
\end{aligned}
\end{equation}
for the nonlocal conductance, in which one assumes that the bias situation is facilitated by $V_L=V, V_R=0$ for local conductance and $V_L=0, V_R=V$ for nonlocal conductance. \\
\indent Having established the conductance calculations, we now turn to current fluctuations. Using the scattering matrix formalism, the current correlation provides a powerful probe of the underlying transport processes in Majorana systems~\cite{Yeyati_1,golub2011shot, pierattelli2025delta, ostrove2021positive, anantram1996current}. In this work, we will mainly focus on zero-temperature, zero-frequency ($\omega=0$) correlators with a finite dc-bias. The current correlation function $S_{LL}$ and the current cross-correlation function $S_{LR}$ \cite{buttiker1992scattering,anantram1996current} can be expressed as:
\begin{widetext}
\begin{eqnarray}\label{SLL}
    S_{LL}(V) \mid_{T \rightarrow 0} &=& \frac{2e^2}{h} \int_0^{eV}  dE  \, \mathbf{Tr} [(\hat{R}_N(E) + \hat{R}_A(E) - (\hat{R}_N(E)-\hat{R}_A(E))^2)] \nonumber \\
&=& \frac{2e^2}{h} \int_0^{eV}  dE  \ \mathbf{Tr} \Bigl[\Bigl (\hat{R}_N(E)(1-\hat{R}_N(E)) + \hat{R}_A(E)(1-\hat{R}_A(E)) + 2\hat{R}_N(E)\hat{R}_A(E) \Bigr)\Bigr],
\end{eqnarray}
\begin{eqnarray}\label{SLR}
S_{LR}(V) \mid_{T \rightarrow 0} &= &-\frac{2e^2}{h} \int_0^{eV} dE  \mathbf{Tr}\Bigl[\left(\hat{R}_N(E)-\hat{R}_A(E)\right) \times \left(\hat{T}_{ECT}(E)-\hat{T}_{CAR}(E)\right) \Bigr].
\end{eqnarray}
\end{widetext}
One must note from the above the important point that the mathematical trace is taken over a product of transmission and reflection operators, unlike in the conductance terms. This is because the current correlators in general have terms that are bi-quadratic in the $s$-matrix elements as noted in App. \ref{shotnoise}, specifically spelled out in \eqref{SLR_final} and \eqref{SLLfinal}. Here, it is clear that the off-diagonal terms in the $s$-matrices are included in the biquadratic combination and the trace taken outside it. Furthermore, unlike current correlations, the sign of cross-correlation $S_{LR}$ is dictated by a competition between two terms, providing a diagnostic of whether the system is dominated by the CPS process ($T_{CAR}$) or the ECT process ($T_{ECT}$). The detailed derivation of these expressions within the scattering matrix framework is provided in App. \ref{shotnoise}. \\
\indent These expressions are then evaluated numerically using the Keldysh non-equilibrium Green's function (NEGF) formalism \cite{Yeyati_1,chevallier2011current,Arora_2024}. We refer to App.~\ref{Keldysh} for the basic details. We start with the $12 \times 12$ matrix form of the MKC Hamiltonian in Nambu space \eqref{eq:Nambu}, $H_{MKC}^{BdG}$ defined as  $\hat{H}_{BdG}= \frac{1}{2} \hat{\Psi}^{\dagger} H^{BdG}_{MKC} \hat{\Psi}$, from which we define the retarded Green's function \begin{equation}\label{eqn:retG}
G^r(E) = \bigl[(E+i\eta) I_{12} - H^{BdG}_{MKC} - \Sigma_L - \Sigma_R\bigr]^{-1}. 
\end{equation}
Here, $\Sigma_{L(R)}$ is the self-energy that describes the coupling to the $L (R)$ contact, which is dependent on the tunneling Hamiltonian that connects the contacts with the MKC. The details of such a calculation \cite{MURALIDHARAN202661} are routine in the limit of dispersionless contacts. 
From the retarded Green's function, we can construct the scattering matrix using the Fisher-Lee relations as explained in App.~\ref{Keldysh}.
The matrix elements mentioned above can then be used to calculate the transmission functions for various processes defined in \eqref{tran}.
The parameter values used for the transport simulations in Fig \ref{fig:Fig3} are outlined in Table \ref{tab:num-params}.
\begin{table}[h]
\caption{Numerical parameters used in all scattering-matrix and NEGF calculations
(identical for both PMM cases). $\Gamma$ is the lead--dot tunnel coupling in
$\Sigma_{L,R} = -i\Gamma/2$; $eV$ is the applied bias for finite-bias noise
(Figs. \ref{fig:Fig3} and \ref{fig:Fig5}); $\eta$ is the regularization in $G^r = [(E+i\eta)\mathbf{1} - H
- \Sigma_L - \Sigma_R]^{-1}$ and $M$ is the number of sites.}
\label{tab:num-params}
\begin{ruledtabular}
\begin{tabular}{lcc}
Parameter & Value & Unit \\
\hline
$\Gamma$ & $5\times 10^{-3}$ & $\Delta$ \\
$eV$     & $0.002$, $0.02$ & $\Delta$ \\
$\eta$   & $10^{-8}$         & $\Delta$ \\
$M$      & $3$               & ---      \\
\end{tabular}
\end{ruledtabular}
\end{table}
\begin{figure}[hbt]
\centering
 \includegraphics[width=1.0\columnwidth]{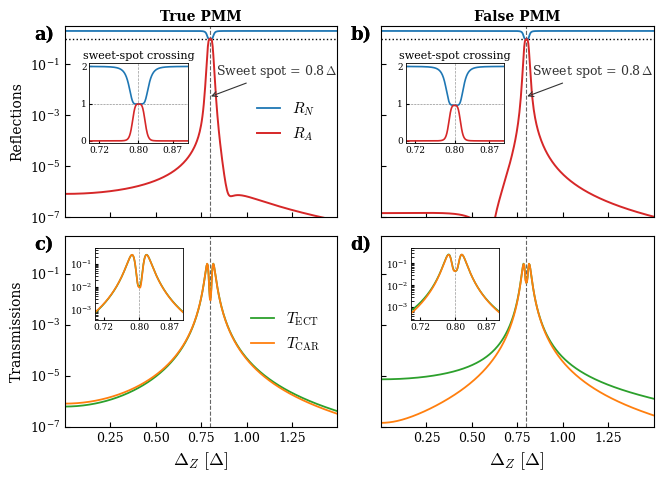}
  \caption{Reflection and transmission calculated at $E=0$, traced over both spin channels, through the device as a function of the Zeeman field, $\Delta_z$. (a), (b) At the sweet spot, the AR reaches its maximum, and the NR reaches its minimum. (c), (d) Away from the sweet spot, the AR probability reduces to less than that of CAR and ECT, in order of magnitude. Both true and false PMMs show very similar primary characteristics around the sweet spot when $\Delta_z$ is tuned.}
\label{fig:Fig2}
\end{figure}
\begin{figure}[hbt]
\centering
 \includegraphics[width=1.0\columnwidth]{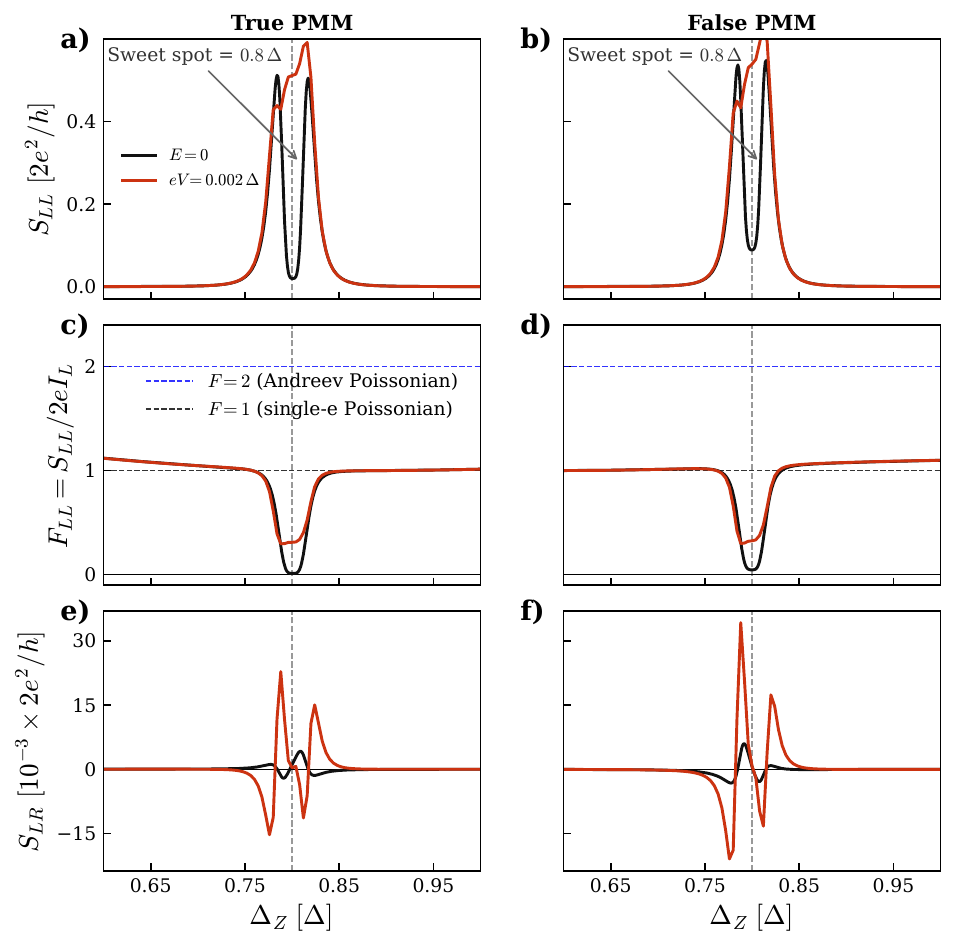}
  \caption{Current correlations and the Fano factor (a) and (b) illustrate the local current correlations $S_{LL}$ at a bias of $eV = 0.002 \Delta$ as a function of the Zeeman field $\Delta_Z$ for true and false PMMs, respectively. We also plot the correlation integrand $S^I_{LL}(E=0)$ to illustrate the cross-term cancellation discussed in the text. (c) and (d) depict the local Fano factor $F_{LL} = S_{LL}/2eI_{LL}$, at negligible bias, which goes to zero at the sweet spot, due to perfect transmission at zero energy. There is also a suppression below the Andreev ($F=2$) limit to the single-electron ($F=1$) Poissonian limit, away from the sweet spot. (e) and (f) show the cross-correlation  $S_{LR}$, with the black line showing the correlation integrand $S_{LR}^I (E=0)$.}
\label{fig:Fig3}
\end{figure}
\begin{figure}[hbt]
\centering
 \includegraphics[width=1.0\columnwidth]{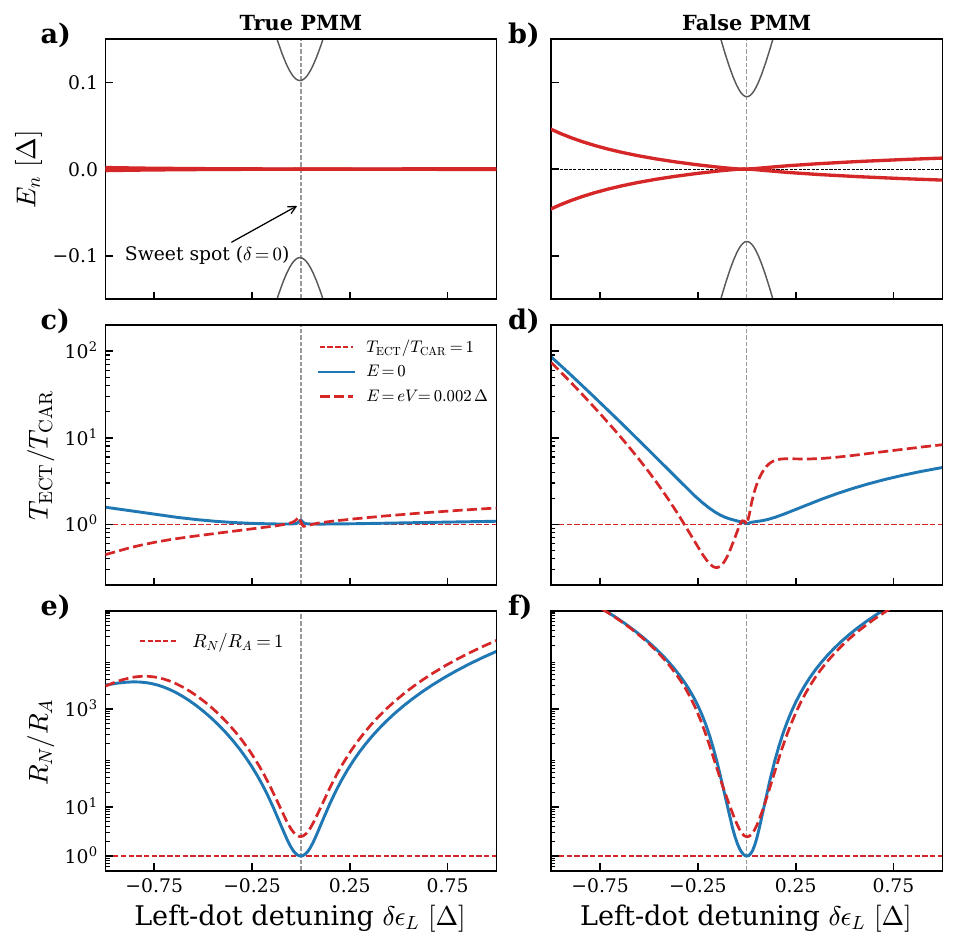}
  \caption{Effect of dot detuning on competing transport processes. (a) and (b) represent the energy spectrum of the system as a function of left-dot detuning $\delta$ for true and false PMM parameters, respectively. (c) and (d) show the corresponding transmission ratio $T_{\mathrm{ECT}}/T_{\mathrm{CAR}}$ at $E=0$ (blue) and at the bias point $E=eV=0.002 \Delta$ (red dash). At the sweet spot ($\delta=0$), both regimes are indistinguishable. However, under finite detuning, the true PMM remains robustly stable, whereas the false PMM is highly sensitive, rapidly losing its zero-energy degeneracy and exhibiting a sharply diverging transmission ratio. (e) and (f) Shows the ratio of $R_N/R_A$ for the same range of detuning at $E=0$ (blue) and at the bias point $E=eV=0.002 \Delta$ (red dash). }
\label{fig:Fig4}
\end{figure}
\begin{figure*}[hbt]
\centering
 \includegraphics[width=2\columnwidth]{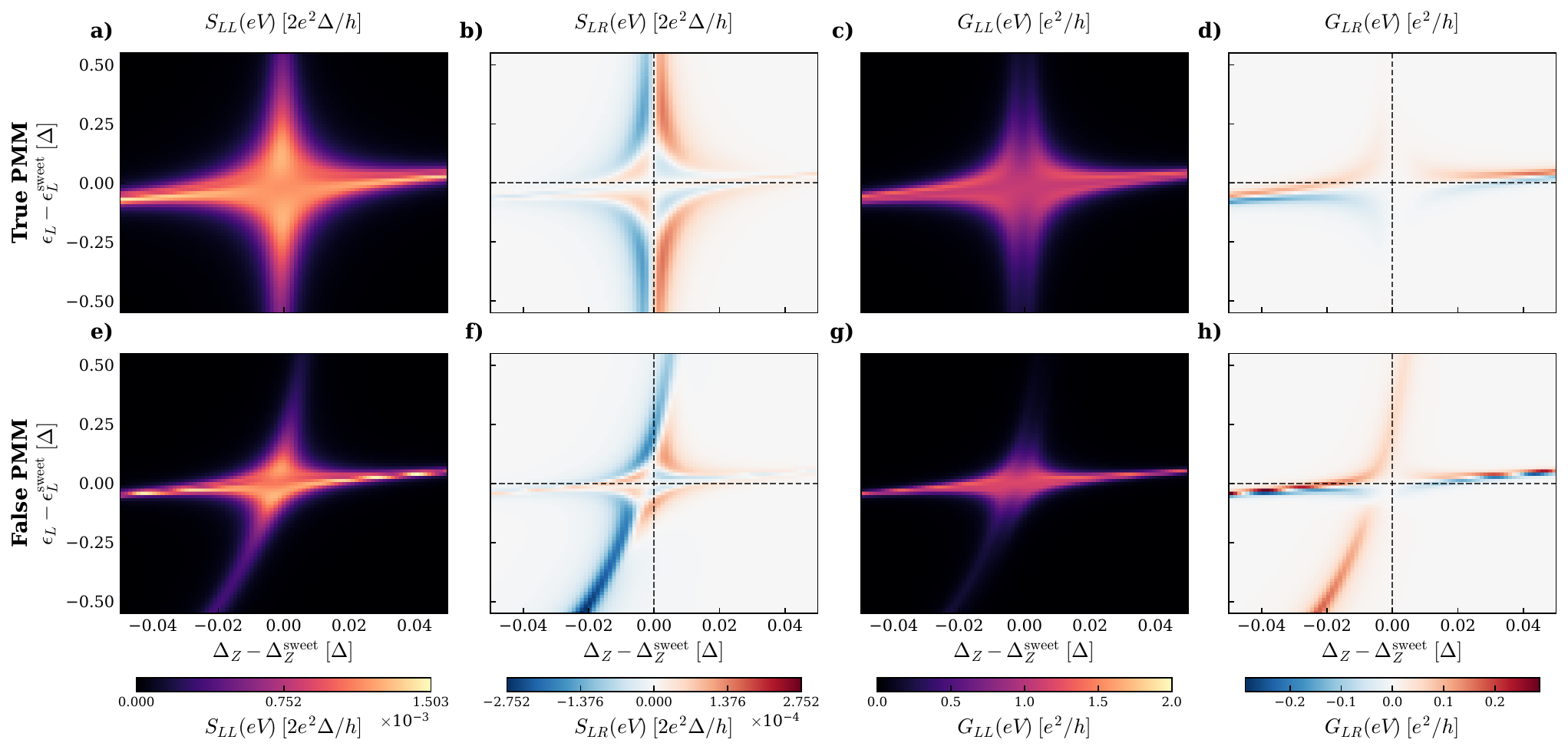}
  \caption{Current correlations and conductance profiles at a small bias, $eV=2 \times 10^{-3} \Delta$. (a)--(d) and (e)--(h) map the transport and noise signatures for true and false PMM configurations, respectively, as a function of Zeeman field detuning $\Delta_Z - \Delta_Z^{\mathrm{sweet}}$ and left-dot chemical potential detuning $\epsilon_L - \epsilon_L^{\mathrm{sweet}}$. The columns display the local correlations $S_{LL}$ [(a), (e)], non-local cross-correlation noise $S_{LR}$ [(b), (f)], local conductance $G_{LL}$ [(c), (g)], and non-local conductance $G_{LR}$ [(d), (h)]. While the local signatures ($S_{LL}$ and $G_{LL}$) remain qualitatively similar near the sweet spot for both cases, the non-local measurements exhibit distinct profiles that help to distinguish between the True and the false PMM states.}
\label{fig:Fig5}
\end{figure*}

\section{Results} \label{sec:results}
\begin{figure*}[hbt]
\centering
 \includegraphics[width=2\columnwidth]{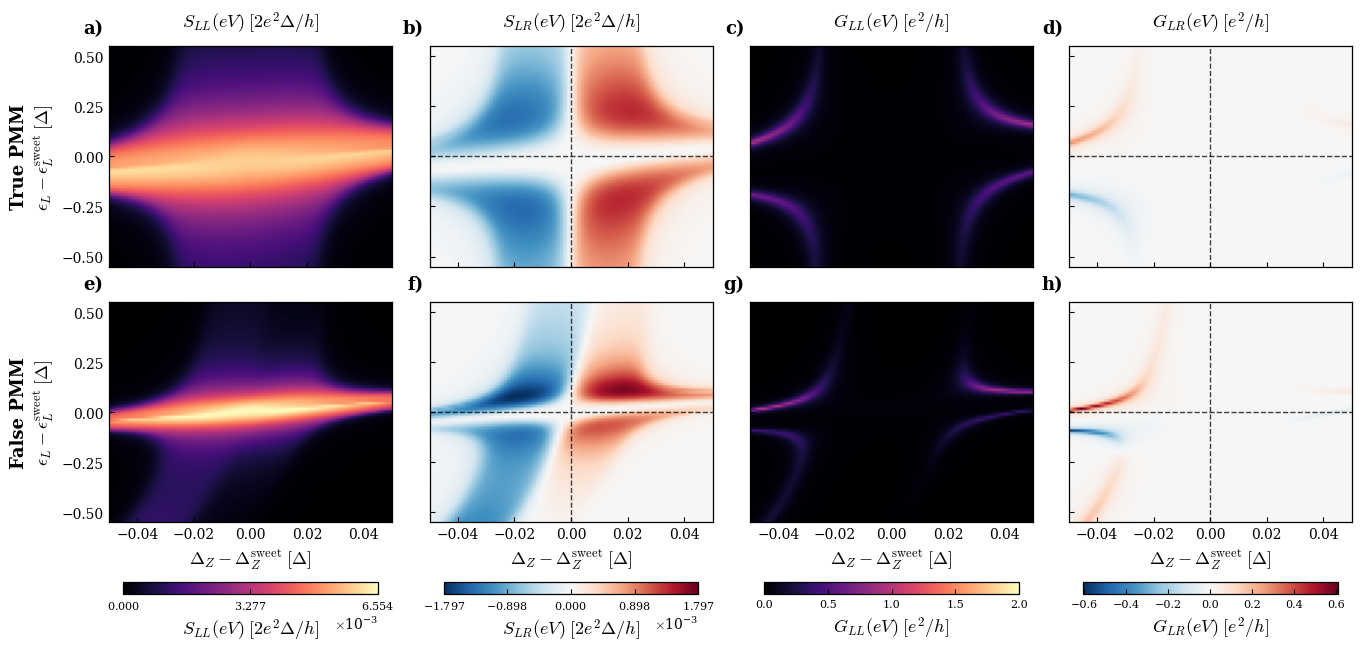}
  \caption{Current correlations and conductance profiles at higher bias $eV=0.02 \Delta$. (a)--(d) and (e)--(h) map the transport and noise signatures for true and false PMM configurations, respectively, as a function of Zeeman field detuning $\Delta_Z - \Delta_Z^{\mathrm{sweet}}$ and left-dot chemical potential detuning $\epsilon_L - \epsilon_L^{\mathrm{sweet}}$. While the conductance signatures, being derivatives at the respective bias, can be very faint, the signatures of current correlations are more pronounced. Specifically, the cross correlations maintain the signature of undergoing a sign change at the sweet spot in the case of true PMMs.}
\label{fig:Fig6}
\end{figure*}
\subsection{Transmission and Reflection}
We first plot various contributions to the overall transmission as a function of the Zeeman field ($\Delta_z$) in Fig~\ref{fig:Fig2}. This gives a detailed picture of the transport processes involved around the fine-tuned sweet spot, which is $\Delta_z$ = 0.8$\Delta$, at $E=0$. More precisely, we plot the trace of the reflection and transmission operators defined in \eqref{tran}, taken over both the up and the down spin channels. We observe the AR and NR values approaching unity at the sweet spot, which physically implies perfect AR for one spin channel and perfect NR for the other. At the sweet spot, the probability of AR is maximum for both true and false PMMs, as depicted in Fig. ~\ref{fig:Fig2}(a) and (b). In this setup, the NR probability is close to unity for both spin channels outside the sweet spots, thus giving a trace of 2, as the quantum dot energies are tuned to create barriers. At the sweet spot, one notices that the AR probability is equal to the NR probability. This situation reflects on the zero-energy states. The contrast with Majorana states in bulk nanowire-superconductor hybrids \cite{Marra,Kejri} is evident, where the probability of AR is unity, since the entire structure is superconducting with all channels undergoing the AR process.  \\
\indent Moving to nonlocal transmission processes, specifically to set the stage for further discussions, in Figs.~\ref{fig:Fig2}(c) and (d) we plot the ECT and CAR transmissions for the true and false PMM cases at $E=0$, respectively. We can notice that at the sweet spot these processes balance each other, and away from the sweet spot the processes are unbalanced. Most importantly, we already notice that for the false PMM, this deviation is more pronounced away from the sweet spot. \\
\indent The aforementioned physics leads to important consequences on the current correlation features shown in Fig.~\ref{fig:Fig3}, with the left (right) panel dedicated to true (false) PMMs. Figs. \ref{fig:Fig3} (a) and (b) show the signatures of the current correlations ($S_{LL}$) for the true and false PMM, respectively. First, let us elaborate on the multichannel physics, wherein one spin channel undergoes AR while the other undergoes NR, and also point to the subtlety related to the biquadratic terms involved in the current correlations defined in \eqref{SLL}. To make both these aspects clear, let us define the integrand $S^I_{ij}(E)$ of the correlator of \eqref{SLL} and \eqref{SLR} as a function of the energy of the scattering state $E$ as
\begin{equation}
S_{ij}(V) \mid_{T \rightarrow 0} = \int_0^{eV} dE S_{ij}^I(E).
\end{equation}
In Fig.~\ref{fig:Fig3}(a) and (b), we plot (black line) $S^I_{LL}(E=0)$ as a function of $\Delta_Z$. The dip around the sweet spot is due to the biquadratic nature of the terms involved in the trace. As noted in the inset of Fig.~\ref{fig:Fig2} (a) and (b), the dip (peak) in $R_N(E=0)$ ($R_A(E=0)$) indicates that one of the channels, up or down spin, has the peak (dip) in the NR (AR) term. If we take the total reflection to evaluate the integrand in \eqref{SLL}, we would get a maximum in $S^I_{LL}(E=0)$. The dip is due to the fact that the trace includes separate channels having a peak (dip) in $R_A$ ($R_N$), where the trace taken after the operation gives the minimum, describing the effect of the bi-quadratic terms \cite{blanter2000shot} involved in the current correlators. \\
\indent It is also evident that local signatures are qualitatively indistinguishable, failing to provide a clear diagnostic of the nonlocal nature of the underlying state at the sweet spot. Similarly, Fig. \ref{fig:Fig3} (c) and (d) display the local Fano factor $F_{LL}=\frac{S_{LL}}{2eI_L}$, respectively, where a significant suppression below the Poissonian limit is observed at the sweet spot ($\Delta_Z \approx 0.8\Delta$). Away from the sweet spot, the Fano factor increases to $F \geq 1$. This transition occurs as the transport becomes primarily governed by single-electron processes (ECT) supplemented by two-electron processes (AR and CAR) that enhance the noise signatures above the single-electron baseline. A Fano factor of zero corresponds to the case of a perfectly transparent normal metal-superconductor junction. Away from the sweet spot, the AR probability reduces by orders of magnitude, the CAR and ECT probabilities become comparable and greater than AR in orders of magnitude, as in Fig~\ref{fig:Fig2}, leading to a Fano factor at the single-electron Poissonian limit of 1 rather than the Andreev limit of two in Fig~\ref{fig:Fig3}. \\
\indent Looking further into this aspect, the Fano factor, in the regions where $R_N \approx 1$ and $R_A \approx 0$, using the unitarity property, the sum-rule and \eqref{SLL}, one can write the local correlations close to zero bias as
\begin{equation}
S_{LL} \approx \frac{4e^3}{h} \left ((T_{ECT}(E=0) + T_{CAR}(E=0) \right) V. 
\end{equation}
With the currents also being written as $I_L=\frac{2e^2}{h} \left ((T_{ECT}(E=0) + T_{CAR}(E=0)\right ) V$, one obtains a Fano factor close to unity. \\
\indent Figures \ref{fig:Fig3} (e) and (f) show the current cross-correlations ($S_{LR}$) for the true and false PMM, respectively. A qualitative difference is visible in the sign-reversal pattern, but more significantly, it is important to note that the cross-correlation goes to zero at the sweet spot. However, the overall magnitudes and structures are otherwise comparable between the two cases, and the sign alone does not provide a sufficiently robust standalone metric to distinguish true from false PMMs. Therefore, it is critical to look for a parameter other than the Zeeman field to detune the system away from the sweet spot, which is why we investigate how the cross-correlation signatures modulate with respect to the detuning of an outer dot. This serves as an additional handle for detuning the system out of the sweet spot and possibly extracting more information through the current cross-correlations. 
\subsection{Effect of dot detuning}
\label{sec:detune}
As seen in Figs. \ref{fig:Fig4} (a) and (b), the near-zero energy levels, highlighted in red, evolve differently for
each type of PMM. The energy spectrum of the false PMM is volatile and is split rapidly as detuning is introduced. In contrast, the true PMM maintains a near-zero energy splitting over a wider range, a behavior indicative of
its role as a precursor to a topologically protected boundary state. This stability is mirrored in the ECT/CAR ratio, which remains more consistent, close to unity, under detuning for the true PMM case, unlike in the false PMM, where the ECT/CAR ratio blows up from unity.\\
\indent Next, by combining the effects of detuning and Zeeman splitting in the density plots shown in Fig.~\ref{fig:Fig5}, it becomes evident that the local conductance [Fig.~\ref{fig:Fig5}(c, g)] and local correlations [Fig.~\ref{fig:Fig5}(a, e)] are virtually indistinguishable between the True and False PMM cases. A defining distinction emerges, however, in the non-local observables regarding the symmetry of the sign-flip; while both cases exhibit a sign inversion around the sweet spot, the geometry of this flip is distinct. In the True PMM, the transition from negative (blue) to positive (red) correlations is confined to a narrow vertical resonance. \\
\indent In contrast, the non-local signals of the false PMM form a prominently diagonal pattern, around $\Delta_Z = \Delta_Z^{sweet}$ where the sign flips. This qualitative difference in the cross-correlation provides a robust experimental fingerprint for identifying the True PMM state, a phenomenon we have consistently observed across other system parameters. This effect is further exaggerated in the case of higher bias, as shown in Fig. ~\ref{fig:Fig6} (a - h). Typically, the current cross-correlation changes signs at the sweet spot even with a modest detuning, as noted by comparing the structures of the maps in Fig. ~\ref{fig:Fig6}(b) and (f). \\
\indent Moving on to nonlocal transport, it is well known that the sign of nonlocal conductance itself can change around the point where $G_{LR}=0$, which occurs both at the beginning of the topological superconductivity regime and at the PMM sweet spot. However, this transition only occurs when $G_{LR}$ is plotted as a function of the bias and is already quite faint. As the bias is ramped up within the superconducting gap, one can notice an absolutely featureless conductance trace by comparing Figs. \ref{fig:Fig5}(c)(d) (g),(h) with Figs. \ref{fig:Fig6}(c)(d) (g),(h). However, current correlations and cross-correlation signatures show a strong variation with bias. This contrast highlights the key advantage of noise-based diagnostics over conventional conductance spectroscopy. The robustness of this signature further suggests that it is not an artifact of a particular fine-tuned parameter set.

\section{Discussion} \label{disc}
Having depicted the cross-correlation signatures clearly, we wish to make a few observations. Firstly, the PMM system within the minimal chain limit is quite different from the bulk topological superconductivity in nanowire-superconductor hybrids. In the latter case, conductance quantization $G_{LL}=\frac{2e^2}{h}$ is a topological consequence of $R_A=1$, which can be directly deduced from \eqref{4}. In addition, in this case $T_{ECT}=T_{CAR}=0$, leading to $G_{LR}=0$, which marks the onset of the topological regime \cite{PhysRevB.97.045421}. In stark contrast, at the PMM sweet spot, as noted in Fig.~\ref{fig:Fig2} (a) and (b), one has $R_N \approx R_A$ with $T_{ECT} = T_{CAR}$, and once again, one can note from \eqref{eq5} that $G_{LR}=0$. \\
\indent However, the change in sign of the current cross-correlation is a robust signature and can become notable as the bias is increased across the parameter range around the sweet spot. This feature, that is, robustness of the signature as one tunes away from the sweet spot, would not be prominent in the conductance signatures. This aspect is evident since the conductance is a differential quantity and only shows peaks or signatures when the average current shows changes, specifically as levels in the energy spectrum appear within the bias window. Given the current experimental efforts on PMMs in minimal chains, the nonlocal signatures yielded these signatures, albeit quite faint in resolution. On the other hand, a more robust signature of nonlocal correlations, whether one is talking about entanglement generation in CPS, or topological superconductivity and connection to topological entanglement entropy or the detection of PMMs, would be what current cross-correlation provides. \\
\indent Toward a plausible experimental realization, ultrasensitive noise measurements are most often performed either using cross-correlation techniques around 1-2 MHz \cite{DiCarlo2006} or utilizing interferometric methods at microwave frequencies \cite{Parmentier2011}. Sometimes cross-correlation of noise is detected at microwave frequencies \cite{Nieminen2016}. Low-frequency noise cross-correlation measurements are made with tuned circuits which effectively convert current fluctuations to voltage noise, yielding a sensitivity of $\sim 1\times 10^{-30}$ $\mathrm{A}^2/\sqrt{\mathrm{Hz}}$ \cite{Bartolomei2020,Lee2023}. Microwave noise experiments may reach almost equally good sensitivities, but longer integration times are required. The problem with microwave systems is that it is hard to properly match the impedance between quantum dots and the 50 \ $\Omega$ environment, and the basic noise temperatures of the amplifiers are inherently larger. \\
\indent The current-current cross correlation in Kitaev chain experiments will be in the range of $\sim 1\times10^{-30}$\; $\mathrm{A}^2/\sqrt{\mathrm{Hz}}$ when using aluminum with a superconducting gap of $\Delta=0.18$\ meV and bias voltage of $0.01 \Delta/e$. This is indeed within the present state of the art, but it would be desirable to work at even smaller bias voltages. In principle, slightly better sensitivities than $\sim 1\times10^{-30}$\; $\mathrm{A}^2/\sqrt{\mathrm{Hz}}$ can be reached by carefully optimizing both the cooled amplifiers and their noise matching at the level of quantum resistance. Furthermore, the cross-correlation signal can be substantially enhanced by using a larger gap superconductor, such as MoRe. However, the low coherence length of MoRe may result in problems in preserving the CAR current in the setup.
\section{Conclusion}
\label{sec:conclusion}
In this work, we investigated a three-site MKC to establish current correlation spectroscopy as a means of going beyond conductance spectroscopy to characterize PMMs. The robustness of the PMM modes and their stability against delocalization, we showed, was embedded in the relative magnitudes of the CAR and ECT processes, which could be extracted via cross-correlation spectroscopy. This effect is adeptly captured by current cross-correlations, whose features showed remarkable stability with respect to the detuning of the quantum dots, making this a prominent feature and a diagnostic for true PMMs even in the short chain limit. We established that current correlation measurements via controlled detuning can provide a concrete framework for unambiguously verifying the topological nature and can also be used to probe entanglement signatures of zero-energy states.
\section*{Acknowledgements} The author BM acknowledges insightful discussions with S. Datta. The author BM acknowledges funding from the Department of Science and Technology (DST), Government of India, under the National Quantum Mission {through Grant no. DST/QTC / NQM/QMD/2024/4} and the Inani Chair Professorship fund {, through Grant No. DO/2024-INAN/001-001}.  The authors BM and AS acknowledge funding from the Dhananjay Joshi Endowment award from IIT Bombay, {through Grant No: DO/2023-DJEF002}. BM also acknowledges funding through the CNRS senior scientist visitor's program at the Center de Physique Théorique UMR 7332, which made this collaboration possible. BM also acknowledges funding through the InstituteQ visitor program, the coordinating organization of the Finnish quantum initiative, which made a collaborative visit to Aalto University possible. The work of PJH was supported by the Jane and Aatos Erkko foundation (SuperC project) and RCF grants no 368333 and no 374170 (Finnish Centre of Excellence in Quantum Materials, QMAT).
The author TM acknowledges the support of the project ``ANY-HALL" (ANR Grant No.
ANR-21-CE30-0064-03), and received support from the French government under the France 2030 investment plan, as part of the Initiative d’Excellence d’Aix-Marseille Université A*MIDEX. TM acknowledges support from institute AMUtech (AMX-19-IET-01X).
\appendix
\section{$s$-Matrix formulation}\label{smatrix}
We first start with the scattering matrix formalism \cite{blanter2000shot, buttiker1992scattering, anantram1996current,Datta_ETMS} to obtain closed form expressions for the currents and current correlation signatures. We define the $s$-matrix elements as $s_{i\alpha, j \beta}$, where, $s_{\alpha \beta}^{ij}$, corresponds to the $s$-matrix element coupling the $\alpha \in e,h$ sector of the $i \in L,R$ to the $\beta \in e,h$ sector of the $j \in L, R$ contact. In order to proceed with the transport calculations, namely, the currents and eventually the current correlators, we start by defining the $A$ matrices \cite{buttiker1992scattering,anantram1996current}, with the same convention as before, i.e., Roman induces for contacts and Greek indices for the electron (hole) sectors:
\begin{equation}
A_{\gamma \delta}^{jl}(i\alpha, E) = \left( \delta_{ij} \delta_{il} \delta_{\alpha \gamma} \delta_{\alpha \delta}- {s^{\dagger}_{i \alpha j \gamma}}(E)  s_{i \alpha l \delta}(E) \right) ,
\label{AMatrix}
\end{equation}
where $\delta_{ij}$ represents the Kronecker delta function, the $s$-matrix elements, $s_{i \alpha, j \beta}$ represent the scattering from the $j \beta$ to the $i \alpha$ sector at a scattering energy $E$. 
\section{Current correlation calculations}\label{shotnoise}
Using the $A$ matrices defined in \ref{AMatrix}, we can further define the zero frequency current operator as
\begin{equation}
\hat I_i(\omega=0)=e\sum_{j,l}
\sum_{\alpha,\gamma,\delta}
\operatorname{sgn}(\alpha)
\int dE\,
A_{\gamma\delta}^{jl}(i\alpha,E)\,
\hat a_{j\gamma}^{\dagger}(E)
\hat a_{l\delta}(E) .
\label{CurrentOperator}
\end{equation}
For the current fluctuations, one is typically interested in the quantity of the form 
\begin{equation} 
\bar{S}_{i j} (t-t') = \frac{1}{2} \left(\langle \delta I_i(t) \delta I_j(t') \rangle + \langle \delta I_j(t') \delta I_i(t) \rangle\right ),
\end{equation}
where the quantities are $\delta I_i(t) = I_i(t) - \langle I_i(t)\rangle$. To evaluate the spectrum of these fluctuations, one has to go to the Fourier domain $I_i(\omega) = \int dt I_i(t) e^{i \omega t} $, and calculate the dc-quantities by setting $\omega =0$. In this context, one writes the current correlation terms in the Fourier domain as follows. 
\begin{equation}
\tilde{\bar{S}}_{ij} (\omega) \delta(\omega + \omega_1) = \frac{1}{2} \left(\langle \delta I_i(\omega) \delta I_j(\omega_1) \rangle + \langle \delta I_j(\omega_1) \delta I_i(\omega) \rangle\right ). \nonumber
\end{equation}
In the above, when $\omega=0$, the RHS is not considered a symmetrized product. Following detailed derivations from B\"uttiker \cite{buttiker1992scattering}, the zero-frequency current correlations can be written as $S_{ij}( \omega = 0) = \langle \delta I_i(\omega=0) \delta I_j(\omega_1=0) \rangle = \int dE S^I_{ij}(E) $. For superconducting hybrid systems, one has to include the electron and hole sectors within the BdG framework. Following the details in Ref. \cite{anantram1996current}, the general form of the current fluctuations can be written as
\begin{equation}
\begin{aligned}
&S_{ij} (\omega =0) = \frac{e^2}{h} \sum_{k,l} \sum_{\alpha,\beta,\gamma,\delta \in \{e,h\}} \operatorname{sgn}(\alpha)\operatorname{sgn}(\beta)
\int dE\\[4pt]
&\quad \quad\mathbf{Tr}\left[A_{\gamma\delta}^{kl}(i\alpha,E)A_{\delta\gamma}^{lk}(j\beta,E)\right]
f_{k\gamma}(E)\left[1-f_{l\delta}(E)\right],
\label{multimode_noise}
\end{aligned}
\end{equation}
where the trace is over spin and transverse-mode indices.
Importantly, here we take the T=0 limit, where the Fermi functions become Heaviside step functions.
At T=0, and for a positive bias applied only to the left lead, the Fermi factors become step functions:
\begin{alignat}{2}
f_{Le}(E) &= \Theta(eV - E),  &\quad f_{Lh}(E) &= \Theta(-eV - E) \nonumber \\
f_{Re}(E) &= \Theta(-E),      &\quad f_{Rh}(E) &= \Theta(-E)
\end{alignat}
The values of the Fermi functions can be tabulated as
\begin{equation}
\begin{array}{c|cccc}
\text{Energy window} & f_{Le} & f_{Lh} & f_{Re} & f_{Rh} \\ \hline
E<-eV      & 1 & 1 & 1 & 1 \\
-eV<E<0    & 1 & 0 & 1 & 1 \\
0<E<eV     & 1 & 0 & 0 & 0 \\
E>eV       & 0 & 0 & 0 & 0
\end{array}
\end{equation}
Outside the interval \(-eV<E<eV\), the product
\(f_{k\gamma}(E)[1-f_{l\delta}(E)]\) vanishes and only the two windows remain
\begin{equation}
-eV<E<0,\qquad 0<E<eV
\end{equation}
which contribute to current correlations.
The full zero-T, zero-frequency noise is then
\begin{align}
S_{ij}
&=
\frac{2e^2}{h}
\int_0^{eV}dE
\sum_{(l,\delta)\in\{(L,h),(R,e),(R,h)\}}
\mathcal{K}_{Le;l\delta}^{ij}(E)
\nonumber\\
&\quad+
\frac{2e^2}{h}
\int_{-eV}^{0}dE
\sum_{(k,\gamma)\in\{(L,e),(R,e),(R,h)\}}
\mathcal{K}_{k\gamma;Lh}^{ij}(E).
\label{two_window_noise}
\end{align}
where we define the kernel;
\begin{equation}
\begin{aligned}
&\mathcal{K}_{k\gamma;l\delta}^{ij}(E) = \sum_{\alpha,\beta\in\{e,h\}}
\operatorname{sgn}(\alpha)\operatorname{sgn}(\beta)\times \\
&\qquad\qquad\qquad\operatorname{Tr}
\left[A_{\gamma\delta}^{kl}(i\alpha,E)A_{\delta\gamma}^{lk}(j\beta,E)\right],
\label{kernel_def}
\end{aligned}
\end{equation}
The two windows in Eq.~\eqref{two_window_noise} are not independent.
Using particle-hole symmetry of the BdG scattering matrix, we can use 
\begin{equation}
s_{i\alpha,j\gamma}(-E)
=
s^{*}_{i\bar{\alpha},j\bar{\gamma}}(E),
\qquad
\bar e=h,\qquad \bar h=e.
\label{phs_smatrix}
\end{equation}
and the noise kernel obeys the following.
\begin{equation}
\mathcal{K}_{k\gamma;l\delta}^{ij}(-E)
=
\mathcal{K}_{l\bar{\delta};k\bar{\gamma}}^{ij}(E).
\label{kernel_phs_identity}
\end{equation}
Using Eq.~\eqref{kernel_phs_identity}, the three negative-energy terms map as
\begin{alignat}{1}
\mathcal{K}_{Le;Lh}^{ij}(-E) = \mathcal{K}_{Le;Lh}^{ij}(E)\\
\mathcal{K}_{Re;Lh}^{ij}(-E) = \mathcal{K}_{Le;Rh}^{ij}(E)\\
\mathcal{K}_{Rh;Lh}^{ij}(-E) = \mathcal{K}_{Le;Re}^{ij}(E)
\end{alignat}
Thus the negative-energy window can be written as:
\begin{equation}
\begin{aligned}
&\int_{-eV}^{0}dE
\sum_{(k,\gamma)\in\{(L,e),(R,e),(R,h)\}}
\mathcal{K}_{k\gamma;Lh}^{ij}(E) \\
&=\int_{0}^{eV}dE
\sum_{(l,\delta)\in\{(L,h),(R,e),(R,h)\}}
\mathcal{K}_{Le;l\delta}^{ij}(E).
\end{aligned}
\end{equation}
Therefore the full \(-eV<E<eV\) integral folds to twice the positive-energy window:
\begin{equation}
S_{ij} = \frac{4e^2}{h} \int_{0}^{eV}dE
\sum_{(l,\delta)\in\{(L,h),(R,e),(R,h)\}}
\mathcal{K}_{Le;l\delta}^{ij}(E).
\label{folded_noise_general}
\end{equation}
Substituting the kernel and the A matrices and simplifying them using the Kronecker delta functions, one can obtain the local and nonlocal current correlations:
\begin{widetext}
\begin{equation}
\begin{aligned}
&S_{LL} = \frac{4e^2}{h} \sum_{\alpha,\beta \in \{e,h\}} \;\sum_{(l,\delta)\in\{(L,h),(R,e),(R,h)\}} \operatorname{sgn}(\alpha)\operatorname{sgn}(\beta) \times \int_0^{eV}dE\;\mathbf{Tr}\left[s^{\dagger}_{L\alpha ,Le}\,s^{}_{L\alpha,l\delta}\,
s^{\dagger}_{L\beta,l\delta}\,s^{}_{L\beta, Le}\right],
\label{SLL_expanded}
\end{aligned}
\end{equation}
\begin{equation}
\begin{aligned}
&S_{LR} = \frac{4e^2}{h} \sum_{\alpha,\beta \in \{e,h\}} \operatorname{sgn}(\alpha)\operatorname{sgn}(\beta) \times \sum_{(l,\delta)\in\{(L,h),(R,e),(R,h)\}}
\int_0^{eV}dE\;\mathbf{Tr}\left[s^{LL\dagger}_{\alpha e}\,s^{Ll}_{\alpha\delta}\,
s^{Rl\dagger}_{\beta\delta}\,s^{RL}_{\beta e}\right],
\label{SLR_expanded}
\end{aligned}
\end{equation}
\end{widetext}
\subsection{Simplifying current correlations}
Using row orthonormality, corresponding to the left lead, of the scattering matrix:
\begin{equation}\label{row_orthonorm}
\sum_{\text{all }(l,\delta)} s^{Ll}_{\alpha\delta}\;s^{Ll\dagger}_{\beta\delta} = \delta_{\alpha\beta}\,\mathbf{1}_2.
\end{equation}
Along similar lines as the calculation in the last section, 
\begin{equation}\label{eq:SLL_surviving}
\begin{aligned}
&\sum_{(l,\delta)\in\{(L,h),(R,e),(R,h)\}}\mathbf{Tr}\left[s^{LL\dagger}_{\alpha e}\,s^{Ll}_{\alpha\delta}\,
s^{Ll\dagger}_{\beta\delta}\,s^{LL}_{\beta e}\right]\\
&= \delta_{\alpha\beta}\;\mathbf{Tr}\left[s^{LL\dagger}_{\alpha e}\,s^{LL}_{\beta e}\right]
- \mathbf{Tr}\left[s^{LL\dagger}_{\alpha e}\,s^{LL}_{\alpha e}\,s^{LL\dagger}_{\beta e}\,s^{LL}_{\beta e}\right]
\end{aligned}
\end{equation}

Again expanding the sums over $\alpha,\beta \in \{e,h\}$, one finally obtains: \\ \\ \\

\begin{widetext}
\begin{equation} \label{SLLfinal}
\begin{aligned}
    &S_{LL} = \frac{2e^2}{h}\int_0^{eV}dE\;\mathbf{Tr}\Biggl[s^{\dagger}_{Le,Le}s^{}_{Le,Le} + s^{\dagger}_{Lh,Le}s^{}_{Lh,Le}- \left(s^{\dagger}_{Le,Le}s^{}_{Le,Le} - s^{\dagger}_{Lh,Le}s^{}_{Lh,Le}\right)^2\Biggr]
\end{aligned}
\end{equation}
\end{widetext}
Similarly this can be expressed in terms of the transmission coefficients as in equation \ref{SLL}.
\subsection{Simplifying current cross-correlations}
\noindent
Writing out the full sum over the trace,
\begin{equation}
\begin{aligned}
&\sum_{\text{all }(l,\delta)} \mathbf{Tr}\left[s^{LL\dagger}_{\alpha e}\,s^{Ll}_{\alpha\delta}\,
s^{Rl\dagger}_{\beta\delta}\,s^{RL}_{\beta e}\right]\\[4pt]
&= \mathbf{Tr}\left[s^{LL\dagger}_{\alpha e}\left(\sum_{\text{all }(l,\delta)}
s^{Ll}_{\alpha\delta}\,s^{Rl\dagger}_{\beta\delta}\right)s^{RL}_{\beta e}\right]
\label{eq:full_sum_factor}
\end{aligned}
\end{equation}

\noindent
Now using row orthogonality of the scattering matrix:
\begin{equation}
\sum_{\text{all }(l,\delta)} s^{Ll}_{\alpha\delta}\,s^{Rl\dagger}_{\beta\delta} = 0
\qquad\forall\;\alpha,\beta.
\end{equation}
\noindent
We can write the partial sum as 

\begin{eqnarray}\label{eq:surviving_collapse}
\begin{aligned}
\sum_{(l,\delta)\in\{(L,h),(R,e),(R,h)\}} \mathbf{Tr}\left[s^{LL\dagger}_{\alpha e}\,s^{Ll}_{\alpha\delta}\,
s^{Rl\dagger}_{\beta\delta}\,s^{RL}_{\beta e}\right] \nonumber \\
= -\mathbf{Tr}\left[s^{LL\dagger}_{\alpha e}\,s^{LL}_{\alpha e}\,
s^{RL\dagger}_{\beta e}\,s^{RL}_{\beta e}\right].
\end{aligned}
\end{eqnarray}

\noindent
Substituting into \ref{SLR_expanded}, one obtains
\begin{equation}\label{SLR_reduced}
\begin{aligned}
&S_{LR} = -\frac{4e^2}{h}\sum_{\alpha,\beta}\operatorname{sgn}(\alpha)\operatorname{sgn}(\beta)\times\\[4pt]
&\qquad\qquad\int_0^{eV}dE\;
\mathbf{Tr}\left[s^{LL\dagger}_{\alpha e}\,s^{LL}_{\alpha e}\,
s^{RL\dagger}_{\beta e}\,s^{RL}_{\beta e}\right]
\end{aligned}
\end{equation}
\noindent
Expanding the sums over $\alpha,\beta \in \{e,h\}$ using $sgn(e)=+1$, $sgn(h)=-1$, one finally obtains  \\ \\ 
\begin{widetext}
\begin{equation}\label{SLR_final}
\begin{aligned}
S_{LR} = -\frac{4e^2}{h}\int_0^{eV}dE\;\mathbf{Tr}\Biggl[
&\left(s^{\dagger}_{Le,Le}s^{}_{Le,Le} - s^{\dagger}_{Lh,Le}s^{}_{Lh,Le}\right)\times\left(s^{\dagger}_{Re,Le}s^{}_{Re,Le} - s^{\dagger}_{Rh,Le}s^{}_{Rh,Le}\right)\Biggr],
\end{aligned}
\end{equation}
\end{widetext}
This can then be expressed in terms of the transmission coefficients, defined in \eqref{tran}, as in \eqref{SLL}.
\section{Connection with the Keldysh formalism}\label{Keldysh}
The expressions formulated previously for the local and nonlocal current correlations are then evaluated numerically using the Keldysh non-equilibrium Green's function (NEGF) formalism \cite{chevallier2011current}. Using the device Hamiltonian and the retarded Green's function \eqref{eqn:retG}, we can construct the scattering matrix from the Fisher-Lee relation:
\begin{eqnarray}\label{eqn:retG}
G^r(E) = \bigl[(E+i\eta) I_{12} - H - \Sigma_L - \Sigma_R\bigr]^{-1} \\
S(E) = -I_{8} + i\sqrt{\Gamma}\, G^r(E) \sqrt{\Gamma}\big|_{\text{lead indices}},
\end{eqnarray}
where $I_{8}$ represents the $8 \times 8$ identity matrix and the $\Gamma$ matrix is given by $\Gamma = i(\Sigma - \Sigma^{\dagger})$. The reduced dimensionality of the matrix is due to the geometry of the setup, which, in our case, corresponds to only the left and right quantum dots, QDL and QDR, connected to the leads. The self-energies $\Sigma_L$ and $\Sigma_R$ encode the effect of the semi-infinite leads on the device and are evaluated using the wide-band limit, wherein the lead density of states is assumed constant over the relevant energy window, so that $\Sigma_{L,R}$ reduces to $-i\Gamma_{L,R}/2$ at the dot sites coupled to the respective lead. This approximation is justified since the superconducting gap and bias voltages, considered here, are small compared to the bandwidth of the normal leads. 
\bibliography{references2}
\end{document}